\documentclass[showpacs,amsmath,amssymb,preprint]{revtex4}
\usepackage{epsfig,dcolumn,bm}

\begin{document}

\title{$N\phi$ state in chiral quark model}

\author{F. Huang$^{1,2,3}$}
\author{Z.Y. Zhang$^2$}
\author{Y.W. Yu$^2$}
\affiliation{\small
$^1$CCAST (World Laboratory), P.O. Box 8730, Beijing 100080, PR China \\
$^2$Institute of High Energy Physics, P.O. Box 918-4, Beijing 100049, PR China\footnote{Mailing address.} \\
$^3$Graduate School of the Chinese Academy of Sciences, Beijing,
PR China}

\begin{abstract}
The structures of $N\phi$ states with spin-parity $J^{p}=3/2^-$
and $J^p=1/2^-$ are dynamically studied in both the chiral SU(3)
quark model and the extended chiral SU(3) quark model by solving a
resonating group method (RGM) equation. The model parameters are
taken from our previous work, which gave a satisfactory
description of the energies of the baryon ground states, the
binding energy of the deuteron, the nucleon-nucleon ($NN$)
scattering phase shifts, and the hyperon-nucleon ($YN$) cross
sections. The channel coupling of $N\phi$ and $\Lambda K^*$ is
considered, and the effect of the tensor force which results in
the mixing of $S$ and $D$ waves is also investigated. The results
show that the $N\phi$ state has an attractive interaction, and in
the extended chiral SU(3) quark model such an attraction plus the
channel coupling effect can consequently make for an $N\phi$
quasi-bound state with several MeV binding energy.
\end{abstract}


\maketitle

\section{Introduction}

The $N\phi$ state has first been studied in Ref. \cite{hygao01},
where the authors followed the idea of Ref. \cite{brodsky90} and
estimated the QCD van der Waals attractive force of the $N\phi$
system. They claimed that the QCD van der Waals attractive force,
mediated by multi-gluon exchanges, can be strong enough to form a
bound $N\phi$ state with a binding energy of about $1.8$ MeV. At
the same time they pointed out that it is possible to search for
such a bound state using the $\phi$ meson below threshold
quasi-free photo-production kinematics experimentally. Using a
simple model, the authors calculated the rate for such
subthreshold quasi-free production process using a realistic
Jefferson Laboratory luminosity and a large acceptance detection
system. They concluded that such an experiment is feasible. To our
way of thinking, it is necessary and desirable to study the
possibility of the $N\phi$ bound state via different theoretical
approaches. The $N\phi$ is a light quark system and the study
based on the constituent quark model seems to be significant and
indispensable. Furthermore, since the $N$ and $\phi$ are two color
singlet hadrons with no common flavor quarks, there is no one
gluon exchange (OGE) interaction between these two clusters, thus
the $N\phi$ system is really a special case to examine the
quark-quark interactions and further the interactions between
these two hadrons.

It is a general consensus that the Quantum Chromodynamics (QCD) is
the underlying theory of the strong interaction. However, as the
non-perturbative QCD (NPQCD) effect is very important for light
quark systems in the low energy region and it is difficult to be
seriously solved, people still need QCD-inspired models to be a
bridge connecting the QCD fundamental theory and the experimental
observables. Among these phenomenological models, the chiral SU(3)
quark model has been quite successful in reproducing the energies
of the baryon ground states, the binding energy of the deuteron,
the nucleon-nucleon ($NN$) scattering phase shifts, and the
hyperon-nucleon ($YN$) cross sections \cite{zyzhang97}. In this
model, the quark-quark interaction containing confinement, OGE and
boson exchanges stemming from scalar and pseudoscalar nonets, and
the short range quark-quark interaction is provided by OGE and
quark exchange effects.

Actually it is still a controversial problem for low-energy hadron
physics whether gluon or Goldstone boson is the proper effective
degree of freedom besides the constituent quark. Glozman and Riska
proposed that the Goldstone boson is the only other proper
effective degree of freedom \cite{glozman96,glozman00}. But Isgur
gave a critique of the boson exchange model and insisted that the
OGE governs the baryon structure \cite{isgur001,isgur002}. Anyway,
it is still a challenging problem in the low-energy hadron physics
whether OGE or vector-meson exchange is the right mechanism or
both of them are important for describing the short-range
quark-quark interaction. Thus the chiral SU(3) quark model has
been extended to include the coupling of the quark and vector
chiral fields. The OGE that plays an important role in the short
range quark-quark interaction in the original chiral SU(3) quark
model is now nearly replaced by the vector meson exchanges. This
model, named the extended chiral SU(3) quark model, has also been
successful in reproducing the the energies of the baryon ground
states, the binding energy of the deuteron, and the
nucleon-nucleon ($NN$) scattering phase shifts \cite{lrdai03}.

Recently, we have extended both the chiral SU(3) quark model and
the extended chiral SU(3) quark model from the study of
baryon-baryon scattering processes to the baryon-meson systems by
solving a resonating group method (RGM) equation
\cite{fhuang05kne,fhuang04kn,fhuang04nkdk,fhuang05lksk,fhuang05dklksk,fhuang05np}.
We found that some results are similar to those given by the
chiral unitary approach study, such as that both the $\Delta K$
system with isospin $I=1$ and the $\Sigma K$ system with $I=1/2$
have quite strong attractions
\cite{fhuang04nkdk,fhuang05lksk,fhuang05dklksk}. In the study of
the $KN$ scattering \cite{fhuang04kn,fhuang04nkdk,fhuang05kne}, we
get a considerable improvement not only on the signs but also on
the magnitudes of the theoretical phase shifts comparing with
other's previous work. We also studied the phase shifts of $\pi K$
\cite{fhuang05kp}, and got reasonable fit with the experiments in
the low energy region. All these achievements encourage us to
investigate more baryon-meson systems by using the same group of
parameters.

In this work, we dynamically study the $N\phi$ interaction in both
the chiral SU(3) quark model and the extended chiral SU(3) quark
model by using the RGM. All the model parameters are taken from
our previous work. The channel coupling of $N\phi$ and $\Lambda
K^*$ is considered, and the effect of the tensor force which makes
for the mixing of $S$ and $D$ waves is also investigated. In the
next section the framework of the chiral SU(3) quark model and the
extended chiral SU(3) quark model are briefly introduced. The
results for the $N\phi$ state are shown in Sec. III, where some
discussion is presented as well. Finally, the summary is given in
Sec. IV.

\section{Formulation}

The chiral SU(3) quark model and the extended chiral SU(3) quark
model has been widely described in the literature
\cite{fhuang04kn,fhuang04nkdk,fhuang05lksk,fhuang05kne,fhuang05dklksk},
and we refer the reader to those works for details. Here we just
give the salient features of these two models.

In these two models, the total Hamiltonian of baryon-meson systems
can be written as
\begin{equation}
H=\sum_{i=1}^{5}T_{i}-T_{G}+\sum_{i<j=1}^{4}V_{ij}+\sum_{i=1}^{4}V_{i\bar
5},
\end{equation}
where $T_G$ is the kinetic energy operator for the center-of-mass
motion, and $V_{ij}$ and $V_{i\bar 5}$ represent the quark-quark
and quark-antiquark interactions, respectively,
\begin{equation}
V_{ij}= V^{OGE}_{ij} + V^{conf}_{ij} + V^{ch}_{ij},
\end{equation}
where $V_{ij}^{OGE}$ is the OGE interaction, and $V_{ij}^{conf}$
is the confinement potential. $V^{ch}_{ij}$ represents the chiral
fields induced effective quark-quark potential. In the chiral
SU(3) quark model, $V^{ch}_{ij}$ includes the scalar boson
exchanges and the pseudoscalar boson exchanges,
\begin{eqnarray}
V^{ch}_{ij} = \sum_{a=0}^8 V_{\sigma_a}({\bm r}_{ij})+\sum_{a=0}^8
V_{\pi_a}({\bm r}_{ij}),
\end{eqnarray}
and in the extended chiral SU(3) quark model, the vector boson
exchanges are also included,
\begin{eqnarray}
V^{ch}_{ij} = \sum_{a=0}^8 V_{\sigma_a}({\bm r}_{ij})+\sum_{a=0}^8
V_{\pi_a}({\bm r}_{ij})+\sum_{a=0}^8 V_{\rho_a}({\bm r}_{ij}).
\end{eqnarray}
Here $\sigma_{0},...,\sigma_{8}$ are the scalar nonet fields,
$\pi_{0},..,\pi_{8}$ the pseudoscalar nonet fields, and
$\rho_{0},..,\rho_{8}$ the vector nonet fields. The expressions of
these potentials can be found in the literature
\cite{fhuang04kn,fhuang04nkdk,fhuang05lksk,fhuang05kne,fhuang05dklksk}.

$V_{i \bar 5}$ in Eq. (1) includes two parts: direct interaction
and annihilation parts,
\begin{equation}
V_{i\bar 5}=V^{dir}_{i\bar 5}+V^{ann}_{i\bar 5},
\end{equation}
with
\begin{equation}
V_{i\bar 5}^{dir}=V_{i\bar 5}^{conf}+V_{i\bar 5}^{OGE}+V_{i\bar
5}^{ch},
\end{equation}
and
\begin{eqnarray}
V_{i\bar{5}}^{ch}=\sum_{j}(-1)^{G_j}V_{i5}^{ch,j}.
\end{eqnarray}
Here $(-1)^{G_j}$ represents the G parity of the $j$th meson. The
$q\bar q$ annihilation interactions, $V_{i\bar 5}^{ann}$, are not
included in this work because they are assumed not to contribute
significantly to a molecular state or a scattering process which
is the subject of our present study.

All the model parameters are taken from our previous work
\cite{fhuang05lksk,fhuang05dklksk}, which gave a satisfactory
description of the energies of the baryon ground states, the
binding energy of the deuteron, and the $NN$ scattering phase
shifts. Here we briefly give the procedure for the parameter
determination. The three initial input parameters, i.e. the
harmonic-oscillator width parameter $b_u$, the up (down) quark
mass $m_{u(d)}$ and the strange quark mass $m_s$, are taken to be
the usual values: $b_u=0.5$ fm for the chiral SU(3) quark model
and $0.45$ fm for the extended chiral SU(3) quark model,
$m_{u(d)}=313$ MeV, and $m_s=470$ MeV. The coupling constant for
scalar and pseudoscalar chiral field coupling, $g_{ch}$, is fixed
by the relation
\begin{eqnarray}
\frac{g^{2}_{ch}}{4\pi} = \left( \frac{3}{5} \right)^{2}
\frac{g^{2}_{NN\pi}}{4\pi} \frac{m^{2}_{u}}{M^{2}_{N}},
\end{eqnarray}
with the empirical value $g^{2}_{NN\pi}/4\pi=13.67$. The coupling
constant for vector coupling of the vector-meson field is taken to
be $g_{chv}=2.351$, the same as used in the $NN$ case
\cite{lrdai03}. The masses of the mesons are taken to be the
experimental values, except for the $\sigma$ meson. The $m_\sigma$
is adjusted to fit the binding energy of the deuteron. The OGE
coupling constants and the strengths of the confinement potential
are fitted by baryon masses and their stability conditions. All
the parameters are tabulated in Table I, where the first set is
for the original chiral SU(3) quark model, the second and third
sets are for the extended chiral SU(3) quark model by taking
$f_{chv}/g_{chv}$ as $0$ and $2/3$, respectively. Here $f_{chv}$
is the coupling constant for tensor coupling of the vector meson
fields.

{\small
\begin{table}[htb]
\caption{\label{para} Model parameters. The meson masses and the
cutoff masses: $m_{\sigma'}=980$ MeV, $m_{\kappa}=980$ MeV,
$m_{\epsilon}=980$ MeV, $m_{\pi}=138$ MeV, $m_K=495$ MeV,
$m_{\eta}=549$ MeV, $m_{\eta'}=957$ MeV, $m_{\rho}=770$ MeV,
$m_{K^*}=892$ MeV, $m_{\omega}=782$ MeV, $m_{\phi}=1020$ MeV, and
$\Lambda=1100$ MeV.}
\begin{center}
\begin{tabular}{cccc}
\hline\hline
  & Chiral SU(3) quark model & \multicolumn{2}{c}{Extended chiral SU(3) quark model}  \\
  &   I   &    II    &    III \\  \cline{3-4}
  &  & $f_{chv}/g_{chv}=0$ & $f_{chv}/g_{chv}=2/3$ \\
\hline
 $b_u$ (fm)  & 0.5 & 0.45 & 0.45 \\
 $m_u$ (MeV) & 313 & 313 & 313 \\
 $m_s$ (MeV) & 470 & 470 & 470 \\
 $g_u^2$     & 0.766 & 0.056 & 0.132 \\
 $g_s^2$     & 0.846 & 0.203 & 0.250 \\
 $g_{ch}$    & 2.621 & 2.621 & 2.621  \\
 $g_{chv}$   &       & 2.351 & 1.973  \\
 $m_\sigma$ (MeV) & 595 & 535 & 547 \\
 $a^c_{uu}$ (MeV/fm$^2$) & 46.6 & 44.5 & 39.1 \\
 $a^c_{us}$ (MeV/fm$^2$) & 58.7 & 79.6 & 69.2 \\
 $a^c_{ss}$ (MeV/fm$^2$) & 99.2 & 163.7 & 142.5 \\
 $a^{c0}_{uu}$ (MeV)  & $-$42.4 & $-$72.3 & $-$62.9 \\
 $a^{c0}_{us}$ (MeV)  & $-$36.2 & $-$87.6 & $-$74.6 \\
 $a^{c0}_{ss}$ (MeV)  & $-$33.8 & $-$108.0 & $-$91.0 \\
\hline\hline
\end{tabular}
\end{center}
\end{table}}

From Table I one can see that for both set II and set III, $g_u^2$
and $g_s^2$ are much smaller than the values of set I. This means
that in the extended chiral SU(3) quark model, the coupling
constants of OGE are greatly reduced when the coupling of quarks
and vector-meson field is considered. Thus the OGE that plays an
important role of the quark-quark short-range interaction in the
original chiral SU(3) quark model is now nearly replaced by the
vector-meson exchange. In other words, the mechanisms of the
quark-quark short-range interactions in these two models are quite
different.

With all parameters determined, the $N\phi$ state can be
dynamically studied in the framework of the RGM, a well
established method for studying the interaction between two
composite particles. The wave function of the $N\phi$ system is of
the form
\begin{eqnarray}
\Psi={\cal A}[{\hat \psi}_N(\bm \xi_1,\bm \xi_2) {\hat
\psi}_\phi(\bm \xi_3) \chi({\bm R}_{N\phi})],
\end{eqnarray}
where ${\bm \xi}_1$ and ${\bm \xi}_2$ are the internal coordinates
for the cluster $N$, and ${\bm \xi}_3$ the internal coordinate for
the cluster $\phi$. ${\bm R}_{N\phi}\equiv {\bm R}_N-{\bm R}_\phi$
is the relative coordinate between the two clusters, $N$ and
$\phi$. The ${\hat \psi}_N$ and ${\hat \psi}_\phi$ are the
antisymmetrized internal cluster wave functions of $N$ and $\phi$,
and $\chi({\bm R}_{N\phi})$ the relative wave function of the two
clusters. The symbol $\cal A$ is the antisymmetrizing operator
defined as
\begin{equation}
{\cal A}\equiv{1-\sum_{i \in N}P_{i4}}\equiv{1-3P_{34}}.
\end{equation}
Expanding unknown $\chi({\bm R}_{N\phi})$ by employing
well-defined basis wave functions, such as Gaussian functions, one
can solve the RGM equation for a bound-state problem or a
scattering one to obtain the binding energy or scattering phase
shifts for the two-cluster systems. The details of solving the RGM
equation can be found in Refs.
\cite{wildermuth77,kamimura77,oka81}.

\section{Results and discussions}

\begin{figure}[htb] \vglue 2.0cm
\epsfig{file=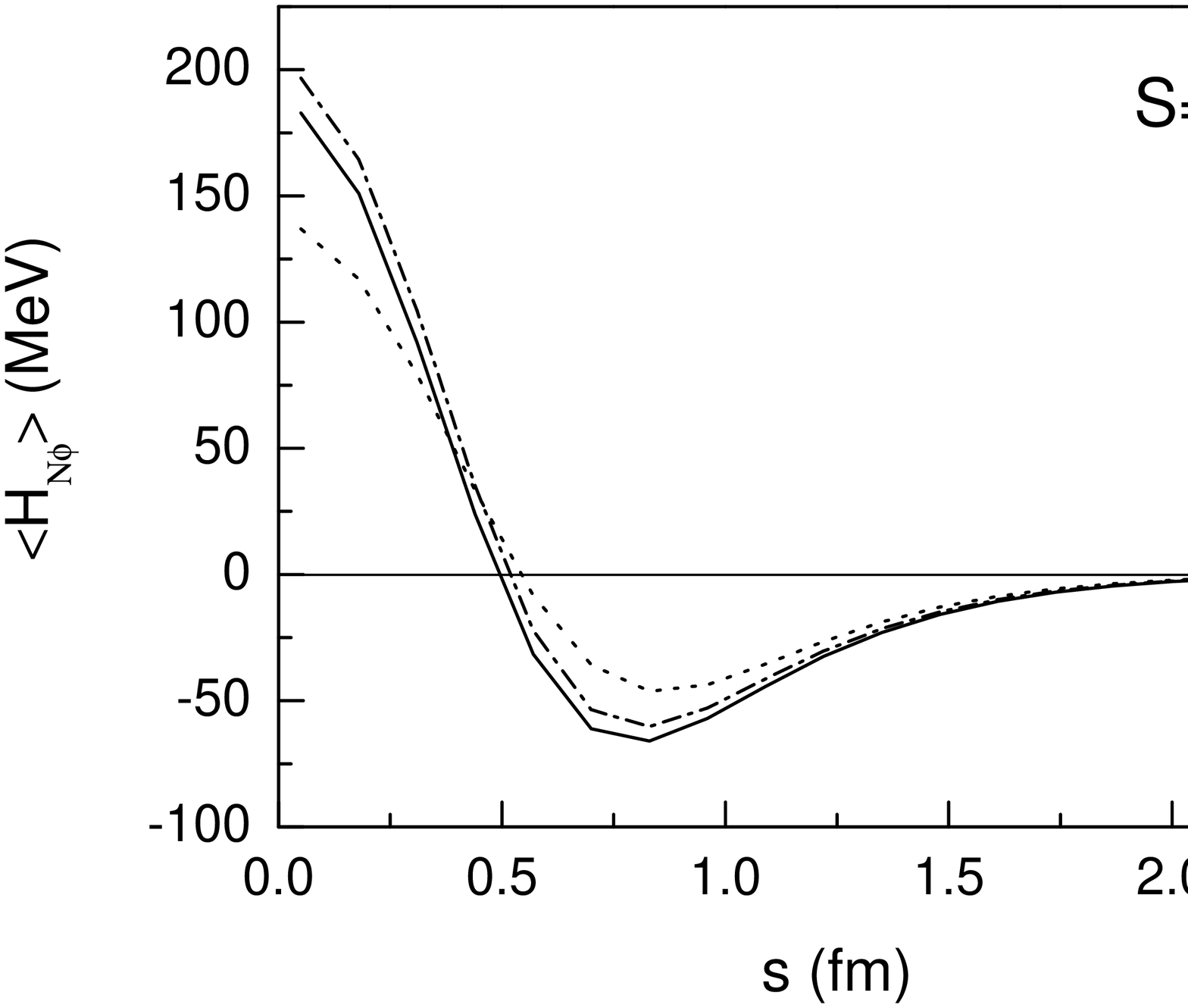,width=7.7cm}
\epsfig{file=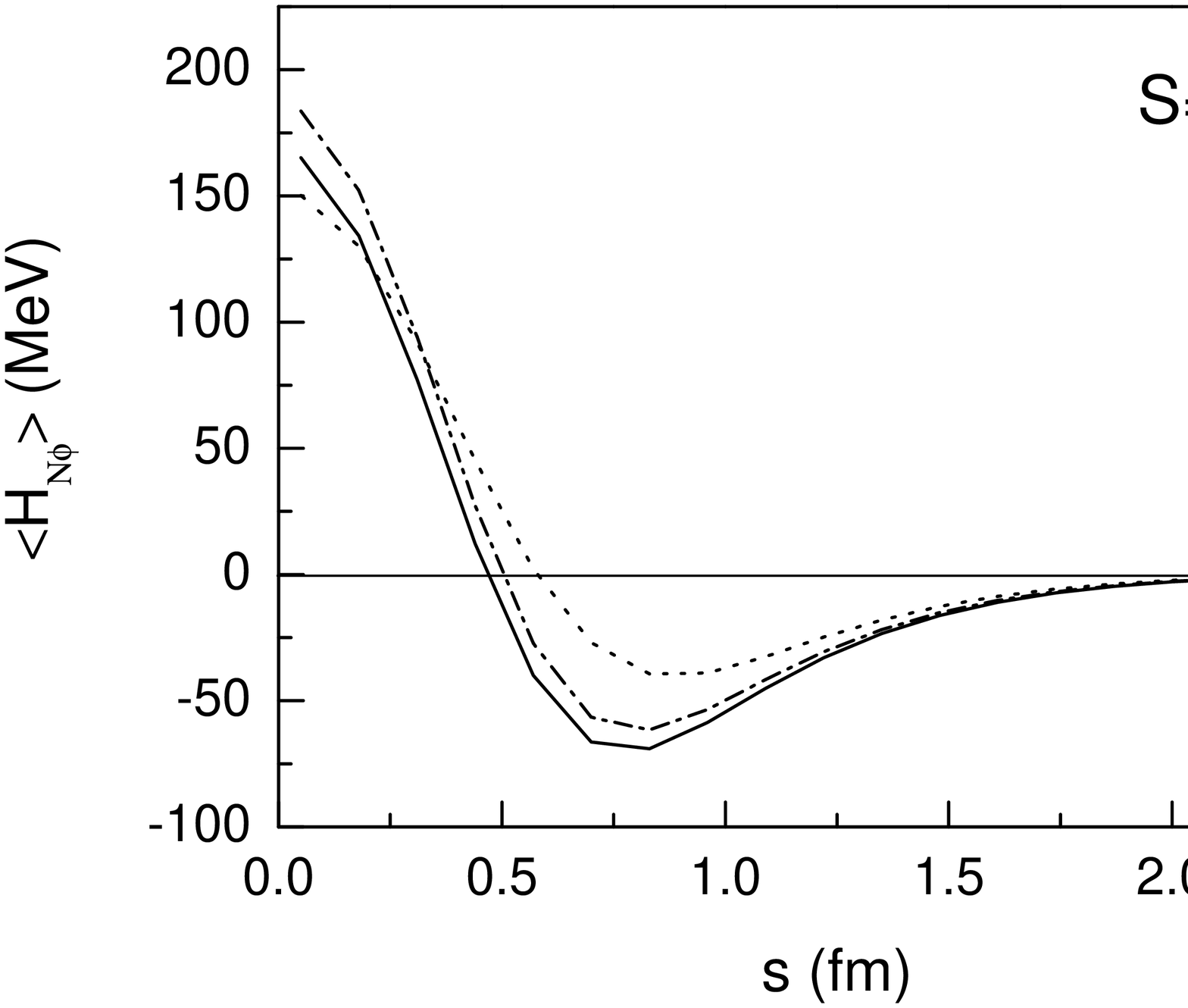,width=7.7cm} \vglue -2.5cm
\caption{\small The GCM matrix elements of the Hamiltonian. The
dotted, solid and dash-dotted lines represent the results obtained
in model I, II and III, respectively.}
\end{figure}

As mentioned above, the $N\phi$ system is a very special
two-hadron state since these two color singlet clusters have no
common flavor quarks. Although the structure of the $N\phi$ state
has already been studied by using the QCD van der Waals attractive
potential in Ref. \cite{hygao01}, a dynamical investigation of
this light quark system in the framework of the constituent quark
model including the coupling of the quark and chiral fields is
still essential. Here we study the $N\phi$ state in our chiral
quark model by treating $N$ and $\phi$ as two clusters and solving
the corresponding RGM equation. Fig. 1 shows the diagonal matrix
elements of the Hamiltonian for the $N\phi$ system in the
generator coordinate method (GCM) \cite{wildermuth77} calculation,
which can be regarded as the effective Hamiltonian of two clusters
$N$ and $\phi$ qualitatively. In Fig. 1, $\langle
H_{N\phi}\rangle$ includes the kinetic energy of the relative
motion and the effective potential between $N$ and $\phi$, and $s$
denotes the generator coordinate which can qualitatively describe
the distance between the two clusters. From Fig. 1, one sees that
the $N\phi$ interaction is attractive in the medium range for both
spin $S=1/2$ and $S=3/2$ cases, this is because this attraction
dominantly comes from the $\sigma$ field coupling and $\sigma$
field is spin and flavor independent. To study if such an
attraction can make for a quasi-bound state of the $N\phi$ system,
we solve the RGM equation for the bound state problem. The results
show that in model I, i.e. the original chiral SU(3) quark model,
and model III, i.e. the extended chiral SU(3) quark model with
$f_{chv}/g_{chv}=2/3$, the $N\phi$ states are unbound for both
spin $S=1/2$ and $S=3/2$, though the $N\phi$ interaction is
attractive. However in model II, i.e. the extended chiral SU(3)
quark model with $f_{chv}/g_{chv}=0$, we get a weakly bound state
of $N\phi$ with about 1 and 3 MeV binding energy for $S=1/2$ and
$S=3/2$, respectively. Actually, as can be seen in Fig. 1, the
$N\phi$ interaction in model II is more attractive than those in
model I and III, thus in model II we can get a weakly $N\phi$
bound state while in model I and III the $N\phi$ is unbound.

{\small
\begin{table}[htb] \caption{The binding energy of $N\phi$ (in MeV).}
\begin{tabular}{ccccccc}
\hline\hline
 Model && \multicolumn{2}{c}{One-channel} &&
\multicolumn{2}{c}{Coupled-channel} \\  \cline{3-4} \cline{6-7}
 && $S=1/2$ & $S=3/2$ && $S=1/2$ & $S=3/2$ \\
\hline
I && $-$ & $-$& & $-$ & $-$ \\
II && 1 & 3 && 3 & 9 \\
III && $-$ & $-$ && 1 & 6 \\
\hline\hline
\end{tabular}
\end{table}}

Since the threshold of $\Lambda K^*$ is only 49 MeV higher than
that of $N\phi$, the channel coupling effect of these two channels
would be un-negligible. This effect is considered by solving a
coupled-channel RGM equation for the bound state problem, and the
calculated binding energies are shown in Table II. One sees that
in model I, the $N\phi$ states are yet unbound for both spin
$S=1/2$ and $S=3/2$ channels, while in model II and III, the
$N\phi$ states are weakly bound with the binding energies of about
3 and 1 MeV for $S=1/2$ and 9 and 6 MeV for $S=3/2$, respectively.
These results tell us that the effect of the channel coupling
between $N \phi$ and $\Lambda K^*$ is considerable and it can make
the $N\phi$ binding energies a little bit larger.

We also study the effect of the tensor force from OGE,
pseudo-scalar and vector field coupling, which results in the
mixing of $S$ and $D$ waves. Our results show that the tensor
force in the $N\phi$ system is very small and its effect can be
neglected. This can be understood easily  because in the $N\phi$
system the tensor force from  OGE and $\pi$ and $\rho$ exchanges
are absent and only $K$, $\eta$, $\eta'$, and $K^*$ exchanges with
the quark exchange can offer tiny tensor force.

As mentioned in Ref. \cite{hygao01}, in the $N\phi$ system, the
OGE is not allowed since the two color-singlet clusters have no
common flavor quarks, and the attraction between $N$ and $\phi$
comes from the QCD van der Waals interaction mediated by
multi-gluon exchanges. In the chiral SU(3) quark model, there is
also no contributions from OGE, while the $\sigma$ exchange
dominantly provides the $N\phi $ attractive interaction. As
regards in the extended chiral SU(3) quark model, there is no
contribution from $\rho$, $\omega$ and $\phi$ exchanges, and the
attraction in this special system also dominantly comes from
$\sigma$ exchange. In our calculation, the model parameters are
fitted by the $NN$ scattering phase shifts, and the mass of
$\sigma$ is adjusted by fitting the deuteron's binding energy,
thus the value of $m_{\sigma}$ is somewhat different for the cases
I, II and III. In model II the mass of the $\sigma$ meson is
smaller than those in model I and III, which means that in model
II $N\phi$ gets more attraction than those in model I and III,
thus much more binding energy of $N\phi$ is obtained in model II
than in model I and III.

Actually, all of the results including the estimation from color
van der Waals force in Ref. \cite{hygao01} are model and parameter
dependent. However while the results obtained from different
theoretical approaches are qualitatively similar, then it would
make sense and this special system becomes more interesting. At
the same time, an experimental measurement of the $N\phi$ binding
energy to examine whether the $N\phi$ system can be bound would be
very important for getting more knowledge of the coupling between
quark and $\sigma$ chiral field.

\section{Summary}

In this work, we dynamically study the $N\phi$ state in the chiral
SU(3) quark model as well as in the extended chiral SU(3) quark
model by solving the RGM equation. All the model parameters are
taken from our previous work, which can give a satisfactory
description of the energies of the baryon ground states, the
binding energy of the deuteron, and the $NN$ scattering phase
shifts. The channel coupling of $N\phi$ and $\Lambda K^*$ is
considered and the effect of the tensor force is also studied. The
calculated results show that the $N\phi$ state has an attractive
interaction, dominantly provided by the $\sigma$ exchange, for
both spin $S=1/2$ and $S=3/2$ channels. The effect of the channel
coupling of $N\phi$ and $\Lambda K^*$ is shown to be considerable,
while the tensor force is displayed to be so small that can be
neglected. In the extended chiral SU(3) quark model, the $N\phi$
attractive interaction plus the coupling to the $\Lambda K^*$
channel can make for an $N\phi$ quasi-bound state with several MeV
binding energy. Experimentally whether their is an $N\phi$
quasi-bound state or resonance state can help us to test the
strength of the coupling of the quark and $\sigma$ chiral field.

\begin{acknowledgements}
This work was supported in part by the National Natural Science
Foundation of China, Grant No. 10475087.
\end{acknowledgements}

\end{document}